\documentclass[journal]{IEEEtran}
\usepackage{graphicx}
\usepackage{paralist}
\usepackage{amsmath}
\usepackage{amssymb}
\usepackage{sidecap}
\usepackage{color}
\usepackage[noadjust]{cite}
\usepackage[table,xcdraw]{xcolor}
\usepackage{tabularx}
\usepackage{soul}
\usepackage{booktabs}
\soulregister\cite7
\soulregister\ref7
\soulregister\textit7
\graphicspath{{./imgs/}}

\ifCLASSINFOpdf
\else
\fi
\begin{document}
	\bstctlcite{IEEEexample:BSTcontrol}
%
% paper title
% can use linebreaks \\ withinto get better formatting as desired
% Do not put math or special symbols in the title.
    \title{Survey of Computer Vision and Machine Learning in Gastrointestinal Endoscopy}
%
%
% author names and IEEE memberships
% note positions of commas and nonbreaking spaces ( ~ ) LaTeX will not break
% a structure at a ~ so this keeps an author's name from being broken across
% two lines.
% use \thanks{} to gain access to the first footnote area
% a separate \thanks must be used for each paragraph as LaTeX2e's \thanks
% was not built to handle multiple paragraphs
%

\author{Anant S. Vemuri
\thanks{}% <-this % stops a space
\thanks{}
\thanks{}
}

\maketitle

% As a general rule, do not put math, special symbols or citations
% in the abstract or keywords.
%\begin{abstract}
%WRITE SOMETHING HERE
%\end{abstract}

% Note that keywords are not normally used for peerreview papers.

% For peer review papers, you can put extra information on the cover
% page as needed:
% \ifCLASSOPTIONpeerreview
% \begin{center} \bfseries EDICS Category: 3-BBND \end{center}
% \fi
%
% For peerreview papers, this IEEEtran command inserts a page break and
% creates the second title. It will be ignored for other modes.
\IEEEpeerreviewmaketitle

This paper attempts to provide the reader a place to begin studying the application of computer vision and machine learning to gastrointestinal (GI) endoscopy. They have been classified into 18 categories. It should be be noted by the reader that this is a review from pre-deep learning era. A lot of deep learning based applications have not been covered in this thesis.

%In this paper thus far, a summary of the various steps involved and the state of the art for scene understanding and image retrieval in computer vision have been presented, with focus on methods that have been encountered in GI endoscopy literature.

\section{Endoscopic Applications}
\label{sec:endoscopic_applications}

The clinical applications have been classified into the following 18 broad categories:

\begin{enumerate} 
	\item \textit{Polyp Detection and Classification (PD):} All colorectal cancers (CRC) develop from dysplastic precursor lesions. This is true either in the presence of a predisposing factor such as in inflammatory bowel diseases (IBD) or lack thereof, with lesions occurring sporadically. Macroscopically the shape of lesions observed in the colon have been classified as described in \cite{Laine2015,Inoue2003}. This class of application involves first detecting the polyps visible during colonoscopy and then presenting a classification based on their type. In \cite{Alexandre2007} feature descriptor using the colour and pixel position in image is used for polyp detection using SVM and in \cite{Alexandre2008}, that feature is compared with Colour Wavelet covariance and LBP. In \cite{Ameling2009a,Ameling2009b}, texture features using GLCM are compared with LBP using SVM for classification. \cite{Angermann2015} proposed using edges, followed by a hough transform of the image before using GLCM for texture features detection. They employed an adaboost classifier. In \cite{Bernal2014c}, firstly a model is defined for polyp appearance as enclosed by intensity valleys, including specular highlights and blood vessels to make the model robust. Using this a polyp localization energy map is generated, which is then used as an input for polyp segmentation. Also refer \cite{Bernal2012b,Bernal2014a,Bernal2015} for more details. \cite{Wang2012}, presented an algorithm, termed as the Classification of Regional Feature (CoRF), that is an extension of the sparse matrix and vector quantization for feature detection and segmentation. CoRF solves the intrinsic block selection problem of vector quantization by including training codebook about the shape of regional feature. They demonstrated that this approach works better for polyp detection and segmentation, as opposed to k-means or LGB clustering. 
	
	In \cite{Kwitt2007}, authors evaluated the discriminative power of image features extracted from sub-bands of the Gabor and the Dual-Tree Complex Wavelet Transform for the classification of zoom-endoscopy images. Further they also incorporated colour channel information and show, that this leads to superior classification results, compared to luminance-only based processing. Later, in \cite{Kwitt2008a}, a colour wavelet cross co-occurrence matrix is proposed and use it to obtain statistical features for classification. This new wavelet-domain based colour texture feature extends the concept of classic co-occurrence matrices to capture information between detail sub-band pairs of different colour channels. The descriptor is then used for poly detection using a KNN classifier with Euclidean distance metric. Further work from the authors can be referred from \cite{Kwitt2008a,Kwitt2008b,Kwitt2007}. In \cite{Gross2012}, an approach to poly classification is presented using vessel segmentation to extract, 22 features that describe the complex vessel topologies. Three feature selection strategies are compared with Simulated Annealing giving the best performance for polyp classification. \cite{Hafner2014c}, proposes a new descriptor by analysing shape of the connected components (blobs). The shape is described using convex hull, skeletonization, perimeter based features and contrast feature histograms for mucosal texture classification of polyps using Pentax iScan chromoendoscopy. The readers are referred to following notable works for further incite; \cite{Ayoub2010,Bae2015,Cheng2011,Engelhardt2010,Eskandari2012,Hsu2012,Hu2015,Iakovidis2005,Iwahori2015,Jadhav2015,Karargyris2011b,Martinez2015,Park2012,Ruano2013b,Stehle2009,Tajbakhsh2014a,Tajbakhsh2013,Tajbakhsh2015b,Tajbakhsh2014c,Takeda2012,Viana2012,Wang2014,Wang2015,Yuan2015a,Yuan2014c,Zhao2011a,Zhao2011b,Zhao2012,Zhou2014b}
	
	\item \textit{Ulcer Detection (UD):} Oesophageal and gastric ulcers are caused as a result of GORD. In the colon ulcerative colitis (a type of IBD) occurs when the lining of the large intestine (colon) and the rectum become inflamed. This inflammation produces tiny sores called ulcers on the lining of the colon. It usually begins in the rectum and spreads upward.
	
	The proposed method in \cite{Charisis2014} involves decomposition of images into components called as intrinsic mode functions, using bi-dimensional ensemble empirical mode decomposition. From the decomposition, two lacunarity based colour texture characteristics were obtained; the second and higher order correlation between intrinsic texture primitives and pixel intensity distribution. An SVM classifier is used on these features for UD. 
	
	In \cite{Charisis2012b}, an overview of the three image decomposition approaches is provided, \begin{inparaenum}
		\item Empirical mode decomposition or EMD ;
		\item Ensemble EMD and;
		\item Bidimensional EEMD
	\end{inparaenum}, that provide the intrinsic mode functions (IMF). A differential Lacunarity (DLac) metric is computed at each IMF and the responses matched with the characteristics of an ulcerated image. Those IMFs that are closely related to the diseased condition are selected for reconstruction of the decomposed image. The DLac response vector computed earlier is used as the feature vector. Using this descriptor, classifier performance comparison between LDA, Quadratic discriminant analysis, NN using Mahalanobis distance and SVM was performed. In \cite{Charisis2010b}, on the other hand, authors use the lacunarity based colour texture features were used to investigate how the structural information of healthy and abnormal tissue is distributed on RGB, HSV and CIE Lab colour spaces.
	
	In \cite{Koshy2015}, an HSV colour space feature histogram was used along with texture features extracted using the Contourlet transform and the Log Gabor filter, which were used to train an SVM classifier for UD. In \cite{Hwang2010}, a curvelet based local binary pattern is proposed for texture feature extraction, to distinguish ulcer from normal regions, by training a multilayer perceptron neural network classifier. Readers are referred to the following references;	\cite{Yuan2015c,Coimbra2006,Eid2013,Karargyris2009,Karargyris2011b,Li2009a,Li2009b,Li2010,LiuD2015,Liu2012,Park2015,Yuan2015c}
	
	\item \textit{Celiac Disease Detection (CED):} Celiac disease is an autoimmune disorder that can occur in genetically predisposed people where the ingestion of gluten leads to damage in the small intestine. During the course of Celiac disease, the mucosa	loses its absorptive villi, leading to a strongly diminished ability to absorb nutrients. The gold standard for detection is based on extraction of biopsies from suspicious regions, which are identified during duodenoscopy using different imaging modalities. Computer aided detection methods to automatically mark suspicious regions during endoscopy have been widely explored in literature to decrease the miss-rates.
	
	\cite{Gadermayr2015c}, presents a CED approach by providing a comparison of classification between LBP, LTP, Multi-Fractal Spectrum, Dual-Tree Complex Wavelet Transform, Shape Curvature Histogram, Fisher vector and Vector of Locally Aggregated descriptors using a linear SVM classifier. They also provide a comparison under NBI, HD zoom endoscopy and standard white light endoscopy. Variants of DT-CWT are explored for automatic classification of endoscopic images using the Marsh classification, in \cite{Uhl2011}. The feature vector was composed of mean and standard deviations of the sub-bands from DT-CWT variant or Weibull parameter of the sub-bands. Enhanced scale invariance was obtained by applying DFT or DCT across the scale dimension of the feature vector. A \emph{k}-NN classifier was used with leave-one-out cross-validation. In \cite{Vecsei2008,Vecsei2009}, spatial domain (histogram) and transform domain (wavelet or Fourier) features are extracted from the images. A comparison between \emph{k}-NN, SVM and bayes classifier is presented. The following references provide further details; \cite{Gadermayr2013a,Gadermayr2013b,Gadermayr2014a,Gadermayr2015d,Gadermayr2015b,Gadermayr2014d,Hammerle2012,Hegenbart2009,Hegenbart2011b,Hegenbart2011a,Hegenbart2012,Hegenbart2013,Uhl2014a,Vecsei2011}.
	
	\item \textit{Crohn's Disease Detection (CRD):} This is another kind of IBD, sometimes attributed to the aggressive immune response to harmless bacteria, by causing inflammation (normal immune system response), leading to chronic inflammation, ulceration, thickening of the intestinal wall, and eventually causing patient symptoms. CRD can occur anywhere from the mouth to the anus but most commonly observed in the ileum and beginning of the colon. \cite{Seshamani2011}, introduces a generic image matching methodology in presence of a complex scene by combining the output of multiple matchers using a single decision function. They provide a study on improving the SVM classifier performance under this framework. \cite{Bejakovic2009} provides the framework for lesion segmentation with application to Crohn's disease.
	
	\item \textit{Haemorrhoid and Bleeding Detection (HD and BD):} Haemorrhoids are itching, painful or bleeding masses of swollen tissues and veins located in the anus and rectum. Bleeding on the other hand could be attributed to wide variety of reasons such as, Angiodysplasia (abnormalities in the blood vessels near the intestines), polyps, ulcers, Crohn's disease, colon cancer, including haemorrhoids. Detection of bleeding thus is very important as it usually indicates a severe condition in the lumen. 
	
	\cite{Abouelenien2013}, use a descriptor comprising of the HSV histogram, dominant colour and texture features from the colour co-occurrence matrix. The dominant colour feature vector included, 8 representative colours, their variances and their percentages in the image. They propose a down-sampling strategy based on unsupervised clustering and probability driven sampling from each cluster to preserve the geometric structure while using fewer instances to train an ensemble of SVM classifiers. \cite{Coimbra2006} present a study of all the MPEG-7 descriptors to determine the ones best suited for BD, UD and PD. Experiments indicated that the best results were obtained when using scalable colour and homogeneous texture descriptors, especially when only relevant coefficients are used using PCA. In \cite{Pan2009,Pan2010}, an ANN classifier was trained using, texture features were extracted in RGB and HSV spaces. An alternate approach was proposed using CIE-LAB colour space with image covariance weighting. \cite{ZhouS2014c} introduced a clipped illumination invariant colour space, to compute an alternate binary feature vector, as opposed to the conventional colour histogram, by comparing similarity between local histograms instead of checking for the existence of a specified pattern. An SVM classifier is trained using this binary feature vector. In \cite{Fu2014}, pixels are grouped through a super-pixel segmentation and, for each super-pixel, the red ratio in RGB space is used as a feature descriptor, which is used to train an SVM classifier. \cite{Ghosh2014} employed the statistical texture descriptors in the hue space to train a \emph{k}-NN classifer. \cite{Ghosh2015} defined an intrinsic colour model using YIO was proposed. This was used to extract statistical features to train an SVM classifier for BD. For further reading, please refer to the following references; HD - \cite{Abe2015,Chen2013} and BD - \cite{Alotaibi2013,Bourbakis2005a,Brzeski2014,Figueiredo2013,Fu2011,Hassan2015,Jung2008,Karargyris2008,Lau2007a,Lee2012,Li2009a,Li2009b,Li2010,LiuJ2009,Liu2012,Mathew2015,Poh2010,Ramaraj2014,Sainju2014,Shah2007,Yuan2015b}
	
	\item \textit{Oesophageal tissue Analysis (OA):} There are two main types of oesophageal cancers; squamous cell cancer and oesophageal adenocarcinoma (OAC). Squamous cell cancer occurs most commonly in people who smoke cigarettes and drink alcohol excessively. Whereas, OAC occurs most commonly in people with gastro-oesophageal reflux disease (GORD). The latter condition has seen an increase in frequency in the last two decades. GORD, a benign complication caused when the stomach acid escapes into the lower part of the oesophagus. As a chronic condition, it leads to changes in the oesophageal lining, causing the tissue to resemble the intestinal wall. This pathological condition is termed as Barrett's oesophagus (BO). Several studies have indicated a direct link of BO with OAC. OAC appears to arise from the Barrett's mucosa through progressive degrees of dysplasia \cite{Conteduca2012,Evans2012} observed in the cells of the lower oesophagus. The possibility of being able to perform staging of the precancerous tissue, provides room for early diagnosis and targeted treatments, avoiding emergency surgical interventions such as oesophagectomy.
	
	%The various chronic conditions occurring in the oesophagus were discussed in \textbf{Section \ref{sec:oesophageal_adenocarcinoma}}. 
	
	The literature reviews methods that include computer-aided detection of these conditions to aide diagnosis. \cite{Huang2015} propose using heterogeneous descriptors computed from heterogeneous colour spaces. Instead of concatenating the descriptors to a super vector, a hierarchical heterogeneous descriptor SVM framework is proposed to simultaneously apply heterogeneous descriptors for GORD diagnosis and overcome the curse of dimensionality problem. \cite{Munzenmayer2009} proposed a content-based image retrieval framework for detection of precancerous lesions in the oesophagus based on colour-texture analysis. The novelty of their approach lies in the interactive loop provided by a relevance feedback algorithm to improve detection accuracy. \cite{Rajan2009}, presented a comparative evaluation of SVM, K-NN and boosting for detection of OA under NBI, WL and chromoendoscopy. \cite{VanDerStap2014a} propose to train an SVM classifier using local colour and texture features, from on the original and on the Gabor-filtered image. Based on the spectral characteristics of the cancerous tissue, specific filters were designed.
	
	\item \textit{Motility Detection (MD):} It is a term used to describe contraction of the muscles that mix and propel contents in the GI tract, with each of the four regions of the GI tract exhibiting specific characteristic movements and are separated by sphincter muscles and abnormal motility or sensitivity in any part of the tract can cause characteristic symptoms \cite{Whitehead2001}. In \cite{Igual2007}, Laplacian of Gaussian filter is used to extract the lumen, then sum of the lumen area throughout the sequence of 9 frames which is compared with two certain thresholds empirically set with the help of the experts. Optical flow based, ego motion estimation is performed and a  Relevance-Vector-Machine classifier is used on the ego-motion representation to extract images with motility. \cite{Segui2013} tackles the problem learning a robust classification function from a very small sample set, when a related but unlabelled data set (for MD) is provided. In \cite{Spyridonos2005}, at the first level of the system, each video was processed resulting in a number of possible contraction sequences. To encode the patterns of intestinal motility, a panel of textural and morphological features of the intestine lumen were extracted. In the second part, the final recognition of contractions sequences was carried out by means of a SVM classifier. \cite{Spyridonos2006}, proposes a novel method based on anisotropic image filtering and efficient statistical classification of contraction features. In particular, the image gradient tensor was applied for mining informative skeletons from the original image and a sequence of descriptors for capturing the characteristic pattern of contractions. \cite{Vilarino2006b,Vilarino2006c} use linear radial patterns by means of the valleys and ridges detection. In this context, they propose descriptors of directional information using steerable filters. Self-organizing maps were used in general summarization for MD. Later, in \cite{Vilarino2010}, use textural, colour and blob features to train a classifier for MD. In \cite{Drozdzal2014b,Drozdzal2013,Drozdzal2015}, propose two sets of features. First, motility based features in which, contractile activity characterization is performed using valley detection through use of Gabor-like filters. Then, the valley image is converted into a 1D signal representing the valley positions. Peak detection is performed that represent contractions in valley positions signal. Second, lumen perimeter estimation is performed, by applying mean-shift clustering to reduce noise in colour distribution. Then on grayscale image, thresholding is performed to segment the lumen. Morphological operators are then used for detection of smooth regions in the intestinal lumen. A combination of, histograms of SIFT Flow Directions to describe the flow course; SIFT descriptors to represent image intestine structure and; SIFT flow magnitude to quantify intestinal deformation, was proposed in \cite{Drozdzal2010}.
	
	\item \textit{Endoscopic Abnormality Detection and Classification (ABD):} This is a broad category that encompasses, all the kinds of lesions or abnormalities that cannot be classified clinically, in any of the above mentioned classes. The methodologies presented here do not focus on any specific disease condition but aim to differentiate a normal tissue from abnormal one. \cite{Ameling2010} provides a review of various feature descriptors used in lesion detection in colonoscopic videos.

	\cite{Barbosa2008} proposed, textural analysis of the different colour channels, using the wavelet transform to select the bands containing the most significant texture information. Later, in \cite{Barbosa2012}, the texture descriptors from co-occurrence matrix at two different scales was used in conjunction with second and higher order moments from the GLCM computed from the image recovered using specific selected scales of the wavelet decomposition of the original image as descriptors. In \cite{Barbosa2009b}, statistical textural descriptors were computed taken from the Discrete Curvelet transform of the image in multiple directions and scales. The covariance of texture descriptors is used as the final feature vector. \cite{Barbosa2010}, performed a comparison between descriptors obtained from wavelet decomposition and discrete curvelet transform. In each case an ANN classifier was trained using the described feature vector. \cite{Francisco2015} proposed using image patches in the BoW model generated using a random forest based clustering which were used to train an SVM classifier. In \cite{Kodogiannis2005,Kodogiannis2007b}, colour histogram statistics were computed for images in R,G,B,H,S,V channels of the WCE images. Additionally, a local texture information was collected for each pixel by using a LTP and labelled as part of a texture unit. This complete information vector is used for classification using a neural network trained using the Bayesian ying-yang method to maximize entropy.

	\cite{Manivannan2015} used image level annotations to learn a set of online local features for adenoma detection in patches extracted in images. The BRISK based spatial structure is used for sampling pixels for learning visual descriptors. \cite{Manivannan2013a}, proposed an extended Gaussian filtered LBP descriptor, robust to illumination changes, noise. The algorithm is claimed to be able to capture more informative edge-like features. \cite{Manivannan2014a} proposes a new method to choose a subset of cluster pairs based on the idea of Latent Semantic Analysis (LSA) and proposes a new inter-cluster statistics which captures richer information than the traditional co-occurrence information. In \cite{Manivannan2013b}, authors present two schemes. The first, working on the full-resolution image, the second on a multi-scale pyramid space. With this framework any feature descriptor could be employed; but a multi-resolution LBP was tested. In \cite{Manivannan2014b}, Root-SIFT and a multi-resolution local patterns descriptors were extracted from image patches, for each colour channel.
	
	For complete set of references, readers are referred to the following list: \cite{Bernal2011a,Chen2011,Coimbra2010,Dhandra2007,Dmitry2014,Hirakawa2013b,Hirakawa2014,Hiroyasu2014,Hu2013,Junzhou2011,Karkanis2003,Lau2007b,Li2012d,Li2010,Li2014b,Li2005b,LiW2011,Liang2012,Lima2009,Lima2012,LiuD2015,Magoulas2004b,Magoulas2010,Majewski2005,Mehlhorn2012,Meng2010,Mewes2011a,Mewes2011b,Mitrea2011,Miyaki2013,Nawarathna2013,Neofytou2006,Prasath2012,Prasath2015,PuertoSouza2015,Riaz2013b,Riaz2011,Riaz2012,Shen2012,Sousa2014,Sun2012,Sun2011,SunZ2012,Surangsrirat2010,Szczypinski2009a,Szczypinski2010,Szczypinski2011,Tjoa2003b,Uhl2014b,Vu2014,WangH2011,Yanagawa2012,Yao2010b,Yuan2013}
	
	\item \textit{Endoscopic Navigation (NAV) and 6-DOF Localization (LOC):} Navigation refers to, using the current endoscopic image information, for determining where to go next. In some ways it charts the path ahead for endoscope. Whereas, localization uses the data from previous few seconds to estimate the current pose or anatomical location of endoscope in the GI tract. This information could be in two forms; as knowledge of the section of GI tract, such as oesophagus, stomach, duodenum, ileum etc., determined by classifying the tissue structure; or secondly, by estimating the complete endoscopic motion to obtain the 6-DOF pose of the endoscope. 
	
	\cite{Armin2015}, modelled the colon as a cylinder. By estimating the camera motion parameters between each consecutive frame, circumferential bands from the cylinder of the colon surface were extracted. Registering these extracted band images from adjacent video frames provided a visibility map, that could reveal unexplored areas by clinicians from colonoscopy videos. \cite{Bell2013} proposed, learning the pose from the optical flow fields in WCE images. Feature descriptors were generated using lumen centred and grid based methodology. ANN was used to evaluate the strength of descriptors extracted from WL and NBI images. In \cite{Iakovidis2013a,Iakovidis2013b}, authors propose extraction of SURF features and use RANSAC based matching to estimate homography between consecutive frames to provide navigational help. \cite{Sfakiotakis2010} proposes to use lumen detection for image-guided visual servoing in endoscopy. For NAV application, the readers are invited to review the following additional references \cite{Bao2014a,Bao2012,Bao2013a,Bao2015,Bao2013b,Bell2014,Burkhardt2014,Chettaoui2006,Khan1996,Kwoh1999,Liu2007a,Liu2013a,LiuJ2015,Mekaouar2009,Reilink2010,Reilink2012,VanDerStap2012,VanDerStap2014b,WangB2014b,WangD2015,Zong2015}.
	
	\cite{Bulat2007,Duda2007}, propose to use, MPEG-7 features along with vector quantization and PCA for descriptor compression. A neural network was trained to classify different section of the GI tract using the computed features. \cite{Figueiredo2015}, propose multi-scale elastic registration of consecutive frames of the WCE and extraction of projective geometry to determine the pose of the capsule endoscope. \cite{Igual2009} employed Gabor filter based texture descriptors to detect duodenum in WCE video stream. For \cite{Liu2009}, the paper proposes a roll angle estimation for complete 6-DOF pose recovery. \cite{LiuL2011} proposes a hybrid tracking method of WCE motion, integrating magnetic sensing and image-based localization. \cite{Mi2014} presents an approach to use the intestinal motility to localize the endoscope. These additional references complete the list for NAV in literature: \cite{Nunez2014,PuertoSouza2014,Spyrou2015,Than2012,Wang2010,Zhou2014a}
	
	\item \label{itm:IAO_IRO_RLoc} \textit{Intra and Inter-Operative Re-localization (IAO and IRO):} The IAO based approaches focus on detecting, tracking and localizing biopsy sites during a single procedure. Primarily, these approaches focus on BSR. One of the first methods in IAO re-localization, was published by Allain \emph{et al.} \cite{Allain2009,Allain2010}. In their approach, the authors proposed to compute feature points in scale-space around the biopsy location and then extracted descriptors for these points using scale invariant feature transform (SIFT) for the two endoscopic views to be matched. Then employing the epipolar constraint, a fundamental matrix was computed between the two views, that mapped the biopsy site to facilitate re-targeting. In \cite{Allain2012} a framework for characterizing and propagation of the uncertainty in the localization of the biopsy points was presented. Mountney \emph{et al.} \cite{Mountney2007} performed a review of various feature descriptors applied to deformable tissue tracking and in \cite{Mountney2006} proposed an Extended Kalman filter (EKF) framework for simultaneous localization and mapping (SLAM) based method for feature tracking in deformable scene, such as in laparoscopic surgery. This EKF framework was then extended in \cite{Mountney2009a} for maintaining a global map of biopsy sites for endoluminal procedures, intra-operatively. The authors presented an evaluation of the EKF-SLAM on phantom models of stomach and oesophagus. Giannarou \emph{et al.} \cite{Giannarou2009} presented an affine-invariant anisotropic region detector robust to soft tissue deformations. This was used by \cite{Atasoy2009} along with SIFT descriptors. The feature matching problem was then modelled as a global optimization of an Markov Random Field (MRF) labelling. Recently, Ye \emph{et al.} \cite{Ye2013,Ye2014} accomplished the biopsy site re-targeting in three stages. First using the Tracking-Learning-Detection (TLD) method proposed by Kalal \emph{et al.} \cite{Kalal2012}. TLD was used for tracking multiple regions around the selected biopsy site. Under the assumption that the regional tissue deformations can be approximated using local affine transformations, a local homography between matched region centres was estimated. In this way multiple regions around the biopsy sites are tracked, which were then used for homography estimation and mapping the biopsy sites. Wang \emph{et al.} \cite{WangB2014a} proposed to learn a graph (atlas) from a sequence of images from several gastroscopic interventions. Considering that the stomach's deformation as not being large between similar frames the nodes of the learnt graph atlas were connected by an estimated rigid transformation. Thus, the mapping of the biopsy sites from a single (reference) frame to subsequent frames for any given intervention was reduced to a graph search problem. Firstly, for the reference frame and the moving frame their corresponding matching nodes in the graph were computed. Using Dijkstra`s algorithm, the shortest path between these matched nodes was obtained. Hence, the transformation between the reference frame and moving frame was obtained as the associated combination of rigid transforms along the shortest path between the corresponding matched nodes of the graph.
	
	In contrast, the IRO methods attempt to provide localization between interventions. \cite{Liu2014} proposed the use of electromagnetic tracking system (EMTS) for localizing the biopsy sites in the stomach. They construct a 3D model of the stomach using SLAM and map the biopsy points tracked using the EMTS on to the 3D model. The inter-operative registration was performed by selecting five reference points manually, during each intervention.
	
	In \cite{Atasoy2011,Atasoy2012}, Atasoy {\em et al.} proposed to formulate the relocalization as a image-manifold learning process. The method involved firstly, building an adjacency graph between the images of a surveillance intervention. Normalized cross-correlation was used as the similarity measure between image frames to compute the adjacency graph. Then using laplacian eigenmaps decomposition that was proposed in \cite{He2005}, a linear projection matrix was computed. This approximation for projection on to the manifold was used to compute the low-dimensional representation for all the images in the intervention. Then, two separate methodologies for performing inter-operative re-localization was proposed using scene association. In \cite{Atasoy2012}, the scene association is performed by computing the nearest neighbour directly over the low-dimensional representation from an earlier surveillance endoscopy. However, in \cite{Atasoy2011} a two-run surveillance endoscopy was suggested, in which a dummy surveillance is performed before, that was used for scene association with the actual surveillance. The authors claimed that the modified approach in \cite{Atasoy2011} allowed for scene association in presence of significant structural changes in the tissue.
	
	For colonoscopic procedures, the need to provide navigational assistance is substantial. One of the earliest approaches involved combination of 3D reconstruction from pre-operative CT with endoscopic video known as virtual colonoscopy. The chief aspect of it involved computation of optical flow to estimate the ego-motion of the colonoscope. Ego-motion or visual odometry involves firstly, extracting features from the image and computing optical flow fields. Then, using the flow fields, the camera motion would be estimated. In \cite{Perto2014} the authors presented a comparison of two ego-motion estimation schemes, supervised and unsupervised. Supervised methods, as shown in \cite{Bell2013} require training data to be available in the form of optical-flow measurements and corresponding camera motion data. Unsupervised approaches, however, used image correspondences between video frames and multiple-view geometry to estimate endoscope motion, as was shown in \cite{Liu2008}. Theoretically, these methods can be applied to oesophageal procedures as well. The first endoscopy can be used to obtain a 3D reconstruction of the oesophagus and can be used in the follow-up surveillance procedures. But, the video based 3D reconstruction in GI procedures is still an open are for research. However, an additional pre-operative imaging such as CT can be used for the reconstruction of the oesophagus. Due to which, such methods were not cost-effective and aren't used as part of routine procedures.

	\item \textit{Lumen Detection (LD):} By itself LD can be employed for NAV, LOC, MD etc. \cite{Asari2000} proposes, global thresholding, followed by a differential region growing using dynamic hill clustering optimization to extract the lumen. In \cite{Gallo2012a}, Haar like feature combined with adaboost were used to select the most discriminative features. Then, a boosted cascade of classifiers was employed for lumen detection. Otsu thresholding was employed for segmenting darker regions of the image in \cite{Kumar1999a}. A pyramidal structure of binarized images was constructed and from the smallest image, the region seed is grown back to the original image resolution to detect the lumen. In \cite{Sanchez2014}, the proposed method is based on the appearance and geometry of the lumen, which we defined as the darkest image region whose centre is a hub of image gradients. In \cite{Tian2001}, the proposed technique applied the Otsu's procedure recursively to obtain a coarse ROI, which is then subjected to an Iris filter operation so that a smaller enhanced region can be identified. The enhanced region was then subjected to the Otsu's procedure recursively and the process of performing Iris filter operation repeated. \cite{Tjoa2003a}, developed a deformable region model approach to extract lumen from the endoscopic image by giving an approximate boundary plan of the lumen using minimum cross-entropy algorithm, that was then deformed to the compute the real boundary automatically.
	
	\item \textit{Uninformative Frame or Region Detection (UI):} Section {sec:challenges} had earlier presented a description of the what constitutes an UI frame. It is important to note in this context that any endoscopic frame need not be completely informative or entirely UI. Thus, some methods proposed in literature also try to identify the UI regions. \cite{Alizadeh2015} propose UI region detection using a multi-stage approach with Chan-Vese segmentation, color range ratio, adaptive gamma correction (AGCM), and finally using canny colour edge detection operator with morphological processing. \cite{An2005}, propose using texture analysis of image DFT and use \emph{k}-means clustering to classify UI frames. \cite{Arnold2009} propose to use L2 norm of DWT decomposition as features given to a Bayesian classifier. In \cite{Bashar2010}, the local colour moments in Ohta space, along with HSV colour histogram were used as features to train an SVM classifier in the first stage of UI frame removal. In the second phase, the Gauss laguerre transform based multi-resolution decomposition was performed and the responses were thresholded. The authors also present a comparison with Gabor and wavelet based descriptors. In the methods proposed by \cite{Bernal2014b}, two values are computed over a grid on the image; \begin{inparaenum}
		\item Dark Region Identification (DRI) using convolution with gaussian kernel.
		\item Directed Gradient Accumulation (DGA)
	\end{inparaenum}. A UI region is then defined by low(DRI) and high(DGA). \cite{Bernal2010} proposed to perform, watershed segmentation followed by morphological closing and Frontier based region merging. After the first merging, region-based merging is performed using mean grey value to threshold over a sliding window. Five empirically chosen region profiles were used for thresholding. In \cite{Fan2011b}, proposed approach involves, lumen detection based on mean shift and evaluation of coherent motility for selecting informative frames. \cite{Ionescu2013}, use texture feature extracted from bank of Gabor filters with a feed-forward neural network for UI classification. \cite{Maghsoudi2014} propose three methods for UI region detection in WCE frames, using feature extracted from morphological operations, statistical features and  Gabor filter based features in HSV colour space. Fuzzy \emph{k}-means, Fisher test and neural network based discriminators were used. The following references give additional methods from this category proposed in literature: \cite{Drozdzal2011,Oh2007,Rangseekajee2011,Rungseekajee2009,Tajbakhsh2014b,WangS2014,WangS2015,Zhou2012}
	
	\item \textit{Specular Highlight Detection and Removal (SHD):} Although, specular highlights in the image constitute UI regions, this particular category of methods attempt to not only identify such regions, but also correct them. In \cite{Arnold2011}, specular highlights is addressed using:a segmentation method based on non-linear filtering and colour image thresholding  followed by a fast in painting method. The proposed method in \cite{Chwyl2015}, aims to decouple the specular and diffuse components of endoscopic imagery in order to suppress specular reflectance. A stochastic Bayesian estimation approach is introduced to estimate the specular component of endoscopic imagery. A Monte-Carlo sampling of image regions is performed for computing posterior probability. \cite{Meslouhi2011}, describe a specularity removal framework using a Dichromatic Reflection Model (DRM) and multi-resolution inpainting technique to obtain the corrected region.
	
	\item \textit{Endoscopic Reconstruction (REC):} 3D Reconstruction in flexible endoscopic procedures is quite a challenging task. Apart from the already discussed, UI frames, presence of repeatable features in a deformable environment poses significant difficulties, if overcome, can aide in assisted diagnosis, pre-operative planning and post-operative review. The feature detectors that were discussed in \ref{Vemuri2016} are used frequently used to recover the 3D from images. 
	
	In \cite{Cai2013}, the tracked feature points are used for estimating camera parameters and providing an estimate of the polyp size. \cite{Castaneda2009} proposed to use SIFT features using normalized SSD based monoSLAM for 3D reconstruction of the oesophagus. \cite{Ciuti2012} employed Shi-Tomasi features and used them in shape from shading framework for reconstruction. \cite{Fan2011a,Fan2010} proposed to use, affine invariant version of SIFT detector and descriptor to estimate the epipolar geometry and recover the 3D. In \cite{Hong2014,Hong2012d}, edges of colon fold contours were first detected and processed to generate the wire frame of the reconstructed virtual colon. A colon fold contour estimation algorithm using a single colonoscopy image was proposed and the depth and shape estimation of colon folds using brightness intensity of pixels was introduced. In \cite{Koulaouzidis2012}, shape-from-shading was used to reconstruct polyps for better recognition. \cite{LiuJ2015} describes an approach to perform a gastric panorama by visual tracking. \cite{WangB2014b} proposes a structure from motion based method that takes advantage of a 6-DOF tracking device that is used to record the endoscope's position during a procedure. After feature tracking, a space constraint strategy is applied to remove the outliers and recover the missing data. In \cite{Zhao2012}, a method to reconstruct the 3D texture surface of the GI tract using single WCE image using Shape from Shading technique is presented. \cite{Zhou2008} used, a circular generalized cylinder as a basis for 3D reconstruction of the GI tract. The model was decomposed as a series of 3D circles and a MRF framework was proposed to maximize the a posteriori estimation. In \cite{Zong2015}, a 3D model and panoramic view are incorporated into the navigation system with three improvements: selection of reference and tracking of features; perspective projection for constructing local and global panoramic view. 3D surface modelling is performed using structure from motion. The system was evaluated for three clinic applications: broadening the endoscopic view, performing non-invasive re-targeting, and determining the overall lesion locations. \cite{Abu2015} proposes to use lumen detection in WCE to create a 3D map using inertial information from the WCE trackers.

	\item \textit{Endoscopic Image Enhancement (IE):} This category refers to a class of approaches directed towards pre-processing steps to improve the quality of visible image and feature response. In \cite{Hafner2013b}, authors evaluate different reconstruction-based super-resolution algorithms in order to enhance the spatial resolution of endoscopic images acquired with an HD endoscope and to determine the its feasibility to study fine mucosal structures in HD endoscopy. To overcome the rather dark WCE images a adaptive contrast diffusion filtering is proposed in \cite{Li2012a}. \cite{Vu2012} proposes an ROI enhancement is for colour correction of regions to be inspected by GI specialists. \cite{Charisis2013} a colour enhancement of WCE frames is proposed to obtain robust texture based features. \cite{Georgieva2015}, proposes use of Homomorphic filtering and \cite{Figueiredo2013} describe an adaptive anisotropic diffusion pre-processing for image enhancement before feature extraction. 
	
	\item \textit{Endoscopic Video Summarization (ES):} This is primarily a category ascribed to wireless capsule endoscopy. Due to the large volume of frames to analyse methods have been developed to minimize this time using different methods. \cite{Zhang2015} proposes a new fast spatio-temporal technique that detects an operation scene a video segment corresponding to a single purpose diagnosis action or a single purpose therapeutic action. In \cite{Zhou2012}, an approach is presented to segment WCE video. To accomplish this, firstly, colour and wavelet texture features are used to denote UI regions. Then boundaries between adjacent organs of WCE video are estimated in two levels. At course level, colour feature is utilized to draw a dissimilarity curve between frames and the aim is to find the peak of the curve, which represents the approximate boundary. At the fine level, Hue-Saturation histogram colour feature in HSV colour space and uniform LBP texture feature from grayscale images are extracted. These features are used to train an SVM classifier for video segmentation. In \cite{Ismail2015}, a two step approach to summarization is proposed. The first step consists of a semi-supervised clustering and Local Scale Learning (SS-LSL) algorithm. This algorithm is used to group video frames into prototypical clusters that summarise the CE video with constraints that are deduced from the training frames. The second step consists of a novel relational motion histogram descriptor that is designed to represent the local motion distribution between two contiguous frames. \cite{Gallo2010} proposes the use of textons for classifying video segments corresponding to different regions in the GI tract. \cite{Fisher2013} reviews various colour and texture descriptor for WCE image analysis. The segments of constant intestinal activity are detected with a robust statistical test that is based on Hoeffding's inequality in \cite{Drozdzal2014a}. \cite{Berens2005} propose using HSV histograms compressed using a combination of DCT and PCA for for identifying different regions in the WCE video. \cite{Yuan2014b} proposes a hierarchical key frame extraction algorithm based on a saliency map to automatically select a small number of key informative frames. \cite{Tsevas2008a} propose a method that is based on clustering using symmetric non-negative matrix factorization, initialized by the fuzzy c-means algorithm and supported by non-negative Lagrangian relaxation, to extract a subset of video scenes containing the most representative frames from an entire examination. \cite{Spyrou2012,Spyrou2013} propose using SURF feature points from consecutive frames, and RANSAC based matching to estimate a homography between consecutive frames for fast video browsing of WCE. An  unsupervised k-window clustering is presented in \cite{Magoulas2004b} to cluster video frames. Each cluster is trained on a different neural network for summarization. \cite{Li2014a} describes a novel colour-texture feature to describe the contents of the frame in a WCE video. Spectral clustering is applied to segment a WCE video into meaningful parts via shot boundary detection using the extracted features. The following references have not been detailed here; \cite{Bourbakis2005b,Cao2007,Cao2006,Chen2015,Chen2009,Dunaeva2014,Fu2012,Gallo2012b,Hai2009,Htwe2011,Huo2012,Hwang2007b,Hwang2008,Lee2013,Lee2007,Li2015}
	
	%\item \textit{Content Based Image Retrieval for Diagnosis or Support (CBIR):} Techniques in this category use computer vision concepts for retrieving similar images from database to aide clinical diagnosis. 
	
	\item \textit{Segmentation of Specific Tissues (IAS):} \cite{Klepaczko2010a} propose a three step approach to segmentation of WCE image frame. \begin{inparaenum}
		\item Local polynomial approximation algorithm which finds locally-adapted neighbourhood of each pixel;
		\item Colour texture analysis which describes each pixel by a vector of numerical attributes that reflect this pixel local neighbourhood characteristics;
		\item Performing k-means clustering based on the colour feature vector.
	\end{inparaenum} For chromoendoscopy and NBI imaging, \cite{Riaz2013a} describes the usage of various visual features individually and in combinations (edgemaps, creaseness, and color), in normalized cuts image segmentation framework. \cite{Suenaga2014} describes an approach to segmenting bubbles in colonoscopic images. 
	
	\item \textit{Clinical Decision Support (CDS):} The methods described in this category discuss approaches to build a generic tool for clinicians to provide decision support. The references described here do not target a specific disease category, but rather a system for detecting lesions and categorizing them to determine the disease type. This category can also be classified as a content based image retrieval platform, which essentially uses computer vision concepts for retrieving similar images from database to aide clinical diagnosis. 
	
	In the method proposed by \cite{Chowdhury2015}, images are transformed to CIE-LAB space. Non-subsampled contourlet transform is used to decompose the chromaticity and intensity components, representing colour and texture features. The decomposed sub-bands are modelled using Generalized Gaussian Density using a ML estimator. The resulting feature vector is then compressed using PCA. Using Least Square-SVM to perform pre-classification, which is followed by computing the kullback-leibler divergence between the features of the query image and the database. \cite{Kalpathy2009}, proposes using GLCM, colour histogram, GIST and Gabor, wavelet, Maximum response (MR8), Leung-Malik (LM) filter bank, and the Schmid Filter banks responses as feature descriptors for image retrieval using a naive Bayes Nearest neighbour classifier. \cite{Khun2009} presents a review of various colour and texture descriptors in to retrieve the types of frames in the endoscopic scene. A comparison is drawn using these feature descriptors by training using ans SVM and an ANN classifier. \cite{Kumar2009}, proposes computing a 10-bin normalized hue and saturation histograms and training a SVM classifier for retrieving the class of the tissue in the ROI. \cite{Kwitt2011} proposes representing the localized features in HD endoscopy images in semantic space to generate a CBIR system for clinicians to review online selected regions. In \cite{Kwitt2010}, using texture features extracted from DT-CWT, the authors propose a generative model based strategy closely related to CBIR for online tissue classification. \cite{Kwitt2012} proposes a novel approach to the design of a semantic, low-dimensional, encoding for endoscopic imagery. \cite{Liu2007b} discusses the development of a platform for image annotation and retrieval in GI endoscopy. A CBIR system is presented in \cite{Munzenmayer2009}, for identifying precancerous lesions in the oesophagus based on color-texture analysis. \cite{Tamaki2013}, explores local features (extracted by using sampling schemes such as Difference-of-Gaussians and grid sampling), BoW, and provides extensive experiments on a variety of technical aspects for feature description. For the CBIR system, an SVM classifier is investigated and its performance under different kernel types, sampling strategies for the local features, the number of classes to be considered etc. is studied. \cite{Yang2013} reviews various feature descriptors and classifications methodologies for WCE images. In \cite{Zhao2011a}, authors propose combining information from multiple images, to design a supervised classification approach using an hidden markov model (HMM) framework. This framework, is prototyped with weak (\emph{k}-NN) classifier to evaluate its performance for regions of the GI tract containing polyps. \cite{Mackiewicz2006} proposes the use of colour features in the form of Hue-Saturation Histograms and texture as SVD of local regions to develop a CBIR system for detecting Pylorus valve between stomach and intestine in WCE. 
	
	In the thesis, \cite{Biswas2014}, from the decomposition of an image using the Hilbert Huang Transform, selected modes were compressed using PCA to generate a representative feature vector. In \cite{Constantinescu2015b}, LBP and its variants were used firstly, for uninformative frame removal and then in the detection of lesions developed due to Celiac disease, Crohn's disease, intestinal polyps and tumours. \cite{Cong2015}, proposes a software system that uses various colour and texture features, combined into a single feature vector. Then a feature selection model is presented, that uses, Deep Sparse SVM (DSSVM) which assigns a suitable weight to the feature dimensions like the other traditional feature selection models and directly excludes useless features from the feature pool. \cite{Boulougoura2004}, presents an intelligent system for online endoscope image analysis. The method discusses extraction of texture features in chromatic and achromatic domains from histograms of each colour component to train an ANN. \cite{DeSousa2008} proposes the use of Hue-Saturation histogram in combination with LBP to classify precancerous and cancerous lesions in multi-spectral imaging. A comparison between Logistical model trees, Naive Bayes, NN and SVM classifiers was made. In \cite{Gadermayr2014c,Gadermayr2015e}, authors study an adaptive texture classification strategy to achieve robustness to varying degrees of degradation in training images. The papers also discuss various similarity measures in this context. \cite{Gadermayr2015a} presents a comparison of various texture based feature descriptors; LBP and variants, multi-fractal spectrum, edge co-occurrence matrix and local phase quantization to train an SVM classifier. Approach to achieve blur invariance through blur-equalization has also been studied in the context of Celiac disease classification. 
	
	In \cite{Hafner2010c}, scale invariant features are extracted from different variants of the DT-CWT of image, in order to classify high-magnification colon endoscopy imagery with respect to the pit pattern scheme. To enhance the scale invariance, the DCT is applied to the feature vectors. The final descriptor contains either consist of the means and standard deviations of the subbands from a DTC-WT variant or of the Weibull parameter of these sub-bands. Readers are referred to review further publications by H\"{a}fner \emph{et al} to study the various use of spatial frequency domain descriptors \cite{Hafner2007a,Hafner2009a,Hafner2009e}. \cite{Hafner2009b} presents a modified version of LBP descriptor over individual colour channels to train a NN classifier using Bhattacharyya distance. \cite{Hafner2012b} presents a comparison of four cross-validation approaches leave-one-image-out, leave-one-parent-image-out, leave-one-lesion-out and leave-one-patient-out for colon polyp classification. \cite{Iakovidis2006} discusses the application areas for decision support in GI endoscopy and presents a review of various features for detection of adenomas in video endoscopy. \cite{Kage2008}, presents a CDS that uses geometrical and colour features from the endoscopic image. The thesis \cite{Kumara2013}, provides an analysis on the detection of various generic scene categories in the colonoscopy videos. \cite{Hafner2010a} discusses usage of edge features and the extraction of most discriminative subsets using a greedy feed forward selection. The descriptors are used with a NN classifier to detect various scenes in GI endoscopy.
	
	The reader is invited to the study following references for further incite; \cite{Hafner2007b,Hafner2011,Hafner2009d,Hafner2010b,Hafner2009f,Hafner2012a,Hafner2014a,Hafner2014b,Hafner2015,Karargyris2010,Keuchel2015,Kumar2012,Kwoh1999,Laranjo2014,Liedlgruber2011a,Liedlgruber2011b,Liu2007b,Mackiewicz2011,Maroulis2003,Mikhaylov2014,Wang2009,Yacob2009,Zheng2005}
\end{enumerate}

This completes the review of scene understanding and classification in GI endoscopy. Although the target application in this thesis has been the oesophagus, a review of the complete GI anatomy was performed since, no such comprehensive review was encountered in literature and a clear understanding of the application domains was felt necessary. Firstly, a description of the various state of the art algorithms was performed, to highlight the key approaches. Then, a classification based on endoscopic application was performed. 18 categories were identified ranging from disease-specific cases, such as CED and CRD, to generalized CDS systems. Application domain of reconstruction, navigation and localization form important parts of an intelligent support systems and hence have also been reviewed. The secondary aim of this categorization was to elucidate the strategic thinking observed in biomedical community, on transfer of technology to GI clinical domain.

\ifCLASSOPTIONcaptionsoff
  \newpage
\fi

% trigger a \newpage just before the given reference
% number - used to balance the columns on the last page
% adjust value as needed - may need to be readjusted if
% the document is modified later
%\IEEEtriggeratref{8}
% The "triggered" command can be changed if desired:
%\IEEEtriggercmd{\enlargethispage{-5in}}

% references section

% can use a bibliography generated by BibTeX as a .bbl file
% BibTeX documentation can be easily obtained at:
% http://www.ctan.org/tex-archive/biblio/bibtex/contrib/doc/
% The IEEEtran BibTeX style support page is at:
% http://www.michaelshell.org/tex/ieeetran/bibtex/
%\bibliographystyle{IEEEtran}
% argument is your BibTeX string definitions and bibliography database(s)
%\bibliography{IEEEabrv,../bib/paper}
%
% <OR> manually copy in the resultant .bbl file
% set second argument of \beginto the number of references
% (used to reserve space for the reference number labels box)
% \begin{thebibliography}{1}
% 
% \bibitem{IEEEhowto:kopka}
% H.~Kopka and P.~W. Daly, \emph{A Guide to \LaTeX}, 3rd~ed.\hskip 1em plus
%   0.5em minus 0.4em\relax Harlow, England: Addison-Wesley, 1999.
% 
% \end{thebibliography}

\bibliographystyle{IEEEtran}
\bibliography{IEEEabrv,References}

\end{document}